# Geoantineutrino Spectrum, $^3$He/$^4$He-ratio Distribution in the Earth's Interior and Slow Nuclear Burning on the Boundary of the Liquid and Solid Phases of the Earth's Core

V.D. Rusov[1,2*], V.N. Pavlovich[3], V.N. Vaschenko[4,5], V.A. Tarasov[1], T.N. Zelentsova[1,2], V.N. Bolshakov[1], D.A. Litvinov[1], S.I. Kosenko[1], O.A. Byegunova[2]

[1] *Odessa National Polytechnic University, Odessa, Ukraine*
[2] *Bielefeld University, Bielefeld, Germany*
[3] *The Institute for Nuclear Researches of National Academy of Science, Kiev, Ukraine*
[4] *Ukrainian National Antarctic Center, Kiev, Ukraine*
[5] *National Taras Shevchenko University, Kiev, Ukraine*

## Abstract

The description problem of geoantineutrino spectrum and reactor antineutrino experimental spectrum in KamLAND, which takes place for antineutrino energy ~ 2.8 MeV, and also the experimental results of the interaction of uranium dioxide and carbide with iron-nickel and silica-alumina melts at high pressure (5-10 GP?) and temperature (1600-2200$^0$ C) have motivated us to consider the possible consequences of the assumption made by V.Anisichkin and coauthors that there is an actinid shell on boundary of liquid and solid phases of the Earth's core. We have shown that the activation of a natural nuclear reactor operating as the solitary waves of nuclear burning in $^{238}$U- and/or $^{232}$Th-medium (in particular, the neutron-fission progressive wave of Feoktistov and/or Teller-Ishikawa-Wood) can be such a physical consequence. The simplified model of the kinetics of accumulation and burnup in U-Pu fuel cycle of Feoktistov is developed. The results of the numerical simulation of neutron-fission wave in two-phase $UO_2$/Fe medium on a surface of the Earth's solid core are presented. The georeactor model of $^3$He origin and the $^3$He/$^4$He-ratio distribution in the Earth's interior is offered. It is shown that the $^3$He/$^4$He ratio distribution can be the natural quantitative criterion of georeactor thermal power. On the basis of O'Nions-Evensen-Hamilton geochemical model of mantle differentiation and the crust growth supplied by actinid shell on the boundary of liquid and solid phases of the Earth's core as a nuclear energy source (georeactor with power of 30 TW), the tentative estimation of geoantineutrino intensity and geoantineutrino spectrum on the Earth surface are given.

_________________________________________________

[*] Corresponding author: Prof. Rusov V.D., E-mail: siiis@te.net.ua

# 1. Introduction

The description problem of geoantineutrino spectrum and reactor antineutrino experimental spectrum in KamLAND [*Araki et al*., 2005], which takes place for antineutrino energy ~ 2.8 MeV, produces a need to consider the probability of the existence of additional energy sources in the interior of the Earth for renewal of geoantineutrino balance. Among such sources there may be the actinids, which are located lower than Gutenberg boundary or, in other words, lower than the mantle. We think that at present the experimental results of Anisichkin et al. [2003; 2005] are the most developed mechanism of actinide shell formation lower than the mantle. According to those results the chemically stable and high-density actinid compounds (particularly carbides and uranium dioxides) almost completely lose their lithophile properties and could be lowered together with melted iron and concentrate in the Earth core due to gravity differentiation of the planet substance. The concentration of actinides on the surface of the Earth's solid inner core could take place after gravity differentiation of substance, i.e. from 4 to $4.5 \cdot 10^9$ years ago. The hypothesis of actinid concentration deep in the planet interior during gravity differentiation of substance was earlier expressed in the works [*Driscoll*, 1988; *Herndon*, 1993; *Anisichkin,* 1997; *Hollenbach and Herndon*, 2001].

The self-propagating waves of nuclear burning in $^{238}$U- and/or $^{232}$Th-mediums must be the natural physical result of existing of such actinide shell in the Earth's core. In other words, in the thermal history of the Earth there must be some geophysical events, which will give a proof of the existence of the spontaneous reactor-like reactions of U-Pu and/or Th-U fuel cycles developed by Feoktistov [1989] and Teller-Ishikava-Wood [1996] respectively on the boundary of liquid and solid phases of the Earth's core. As it is shown below, such geophysical events might be the anomalous $^3$H/$^4$H-ratio distributions in the Earth's interior.

The main purpose of the present paper is trial estimation of the intensity of oscillation geoantineutrino flow on the Earth's surface from different radioactive sources ($^{238}$U, $^{232}$Th and $^{40}$K) by analysis of time evolution of radiogenic heat-evolution power of the Earth within the framework of the geochemical model of the mantle differentiation and the crust growth [*O'Nions et al*., 1979; *Rusov et al*., 2003], which is supplemented by the nuclear energy source located on the boundary of liquid and solid phases of the Earth's core.

# 2. The simulation of Feoktisov's neutron-fission wave

The mechanism of uranium concentration in the Earth core is in detail considered in the work [*Anisichkin et al.*, 2003; *Anisichkin et al*., 2005]. The results of the experiments [*Anisichkin et* *al.*, 2003; *Anisichkin et al*., 2005] on the interaction of uranium carbide and dioxide with nickel-iron and silica-alumina melts at high pressure (5÷10 GPa) and temperature (1600÷2200° ?) give grounds to consider that on the early stages of the evolution of the Earth and other planets uranium and thorium oxides and carbides (as the most dense, refractory and marginally soluble at high pressures) could accumulate from the magma "ocean" on the solid inner core of the planet, thereby creating the possibility for the activation of chain nuclear reactions [*Anisichkin et al.*, 2003; *Anisichkin et al*., 2005] and, particularly of Feoktisov [1989] and/or Teller-Ishikawa-Wood [1996] mechanism of progressing wave.

The geometric image of the natural "stationary" fast reactor, according to the work [*Feoktistov*, 1989], could be pictured in the following way. Consider an infinite cylinder of $^{238}$U about 1m in diameter. In some part of it there is a reaction focus formed forcedly, for example, due to enrichment by fissionable isotope. The next layers of uranium catch the neutrons escaping from reaction area and then $^{239}$Pu is efficiently produced in these layers. If the energy-release is sufficiently high, the concentration of $^{239}$Pu in adjoining areas becomes greater than the critical one and center of energy-release will shift. At the same time the accumulation of plutonium in next layers will begin. So, as result of such a fuel cycle (first proposed by Feoktistov in 1989).

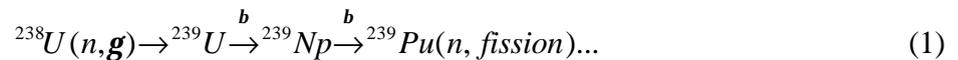

$$^{238}U(n,g) \rightarrow ^{239}U \overset{b}{\rightarrow} ^{239}Np \overset{b}{\rightarrow} ^{239}Pu(n, fission)... \qquad (1)$$

a progressing wave will arise, on front of which uranium is transformed to plutonium due to fission neutrons. In other words, neutron-fission wave transmission in $^{238}$U-medium is possible at a certain correlation between the equilibrium ($n_{Pu}$) and critical ($n_{crit}$) concentrations of plutonium, i.e. ($n_{crit} < n_{Pu}$). A wave velocity is about $L/t \sim 1.5$ cm/day (where $L \sim 5$ cm is diffusion distance of neutron in uranium and $t = 2.3/ln2 = 3.3$ days is time of plutonium formation by $\beta$-decay of $^{239}$U). Note that besides delay time of neutrons one more time $t_{1/2} = 2.3$ days (which plays an important role in safety of Feoktistov natural reactor [*Feoktistov*, 1989]) appears in scheme (1).

The similar idea underlies the mechanism of the formation of nuclear burning progressing wave in $^{232}$Th-medium corresponding to Teller-Ishikawa-Wood Th–U fuel cycle

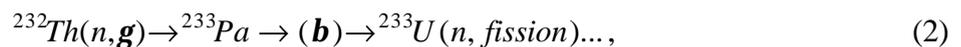

$$^{232}Th(n,g) \rightarrow ^{233}Pa \rightarrow (b) \rightarrow ^{233}U(n, fission)..., \qquad (2)$$

which was described in 1996 in the work [*Teller et al*., 1996].

In our paper the simplified model of Pu accumulation and U burnup kinetics is developed. In this model one-dimensional semi-infinite U-Pu medium irradiated from butt-end by external neutron source is considered in diffusion one-group approximation (neutron energy is ~ 1 MeV). The respective system of differential equations, which describes the kinetics of Feoktistov U-Pu fuel cycle taking into account delayed neutrons, i.e. the kinetics of initiation and transmission of neutron-fission wave $n(x, t)$, looks like:

$$\frac{\partial n(x,t)}{\partial t} = D \Delta n(x,t) + q(x,t), \qquad (3)$$

where

$$q(x,t) = \left[\boldsymbol{n}(1-p) - 1\right] \cdot n(x,t) \cdot \boldsymbol{u}_n \cdot \boldsymbol{s}_f^{Pu} \cdot N_{Pu}(x,t) + \sum_{i=1}^{6} \frac{\widetilde{N}_i \ln 2}{T_{1/2}^i} -$$

$$- n(x,t) \cdot \boldsymbol{u}_n \cdot \left[ \sum_{8,9,Pu} \boldsymbol{s}_a^i \cdot N_i(x,t) + \sum_{i=1}^{6} \boldsymbol{s}_a^i \cdot \widetilde{N}_i(x,t) + \sum_{i=fragments} \boldsymbol{s}_a^i \cdot \overline{N}_i(x,t) \right],$$

$$\frac{\partial N_8(x,t)}{\partial t} = -\boldsymbol{u}_n \cdot n(x,t) \cdot \boldsymbol{s}_a^8 \cdot N_8(x,t), \qquad (4)$$

$$\frac{\partial N_9(x,t)}{\partial t} = \boldsymbol{u}_n \cdot n(x,t) \cdot \boldsymbol{s}_a^8 \cdot N_8(x,t) - \frac{1}{\boldsymbol{t}_b} N_9(x,t), \qquad (5)$$

$$\frac{\partial N_{Pu}(x,t)}{\partial t} = \frac{1}{\boldsymbol{t}_b} N_9(x,t) - \boldsymbol{u}_n \cdot n(x,t) \cdot \left(\boldsymbol{s}_a^{Pu} + \boldsymbol{s}_f^{Pu}\right) \cdot N_{Pu}(x,t), \qquad (6)$$

$$\frac{\partial \widetilde{N}_i}{\partial t} = p_i \cdot \boldsymbol{u}_n \cdot n(x,t) \cdot \boldsymbol{s}_f^{Pu} \cdot N_{Pu}(x,t) - \frac{\ln 2 \cdot \widetilde{N}_i}{T_{1/2}^i}, \quad i = 1,...,6 \qquad (7)$$

To set the last item in the right side of $q(x, t)$ the approach of effective additional neutron absorber was used:

$$n(x,t) \cdot \boldsymbol{u}_n \cdot \sum_{i=fragments} \boldsymbol{s}_a^i \cdot \overline{N}_i(x,t) = n(x,t) \cdot \boldsymbol{u}_n \cdot \boldsymbol{s}_a^{eff} \cdot \overline{N}(x,t), \qquad (8)$$

where kinetic equation for $\overline{N}(x,t)$ taking into account the fact that fission with two fragment formation is most probable has a form:

$$\frac{\partial \overline{N}(x,t)}{\partial t} = 2\left(1 - \sum_{i=1}^{6} p_i\right) \cdot n(x,t) \cdot \boldsymbol{u}_n \cdot \boldsymbol{s}_f^{Pu} \cdot N_{Pu}(x,t) + \sum_{i=1}^{6} \frac{\widetilde{N}_i \ln 2}{T_{1/2}^i}. \qquad (9)$$

Here $n(x,t)$ is neutron density; $D$ is diffusion constant of neutrons; $\boldsymbol{u}_n$ is neutron velocity ($E_n = 1$ MeV, one-group approximation); $\widetilde{N}_i$ are the concentrations of neutron-rich fission fragments of $^{239}Pu$ nuclei; $N_8$, $N_9$, $N_{Pu}$ are $^{238}U$, $^{239}U$, $^{239}Pu$ concentrations respectively; $\overline{N}_i$ are concentrations of the rest fission fragments of $^{239}Pu$ nuclei; $\boldsymbol{s}_a$ is neutron-capture micro-cross-section; $\boldsymbol{s}_f$ is fission micro-cross-section; $t_b$ is nucleus life time in respect of $b$−decay; $p_i$ ($p = \sum_{i=1}^{6} p_i$) are the parameters characterizing the delayed neutrons groups for main fuel fissionable nuclides [*Smelov*, 1978].

The boundary conditions for the system of differential equations (3)-(7) are

$$n(x,t)\big|_{x=0} = \Phi_0 / u_n, \quad n(x,t)\big|_{x=l} = 0, \tag{10}$$

where $\Phi_0$ is neutron density of the plane diffusion source of neutrons located on the boundary $x=0$; $l$ is the uranium block length.

The estimation of neutron flux density of an inner source on the boundary $F_0$ can be obtained by reasoning from the estimation of plutonium critical concentration ~ 10% :

$$4t_b F_0 s_a^8 N_8(x,t)\big|_{t=0} = 0.1 N_8(x,t)\big|_{t=0}$$

and so

$$F_0 \approx 0.1 / 4 t_b s_a^8. \tag{11}$$

Here it should be noted that Eq. (11) is only the estimation of $F_0$. The results of computational experiment show that it can be substantial smaller in reality.

Generally speaking, the different boundary conditions can be used depending on the physical conditions of nuclear fuel neutron "firing", for example, the Dirichlet condition of Eq. (10) type, Neumann condition or so-called third-kind boundary condition, which summarizes first two conditions. The use of a third-kind boundary condition is recommended in the neutron transport theory [*Smelov*, 1978]. This condition, which in the simple case (known as Milne problem) is the linear combination of neutron concentration $n(x, t)$ and its spatial derivative $\partial n/\partial x(x,t)$ on the boundary, looks like

$$n(0,t) - 0.7104 l n^{(1,0)}(0,t) = 0 \tag{12}$$

where λ is the range of neutrons and $n^{(1,0)}(0, t) \equiv \partial n/\partial x\,(0, t)$.

Although the behavior of "neutron source-nuclear fuel" system is different on the boundary depending from the different boundary conditions, but as computational experiment shows the behavior of the system in the active zone (i.e., far from the boundary) is asymptotically invariant. This confirms that the independence of wave propagation in reactor volume on the firing and boundary conditions. In this sense the problem of determining the optimum parameters of nuclear fuel "firing" in the "neutron source-nuclear fuel" system is nontrivial and extraordinarily vital issue, which requires the individual consideration.

The initial conditions for the system of differential equations (3)-(7) are

$$n(x,t)\big|_{x,t=0} = \Phi_0 \big/ \boldsymbol{J}_n, \quad n(x,t)\big|_{x,t=0} = 0; \tag{13}$$

$$N_8(x,t)\bigg|_{t=0} = \frac{\boldsymbol{r}_8}{\boldsymbol{m}_8} N_A \approx \frac{19}{238} N_A, \tag{14}$$

$$N_9(x,t)\big|_{t=0} = 0, \quad N_{Pu}(x,t)\big|_{t=0} = 0, \quad \tilde{N}_i(x,t)\big|_{t=0} = 0, \quad \overline{N}(x,t)\big|_{t=0} = 0, \tag{15}$$

where $\boldsymbol{r}_8$ - is the density, which is expressed in units of g·cm$^{-3}$; $N_A$ - Avogadro constant.

The following values of constants were used for simulation:

$$\boldsymbol{s}_f^{Pu} = 2.0 \cdot 10^{-24} \, cm^2; \; \boldsymbol{s}_f^8 = 0.55 \cdot 10^{-24} \, ?m^2; \tag{16}$$

$$\boldsymbol{s}_a^8 = \boldsymbol{s}_a^i = \boldsymbol{s}_a^{fragments} = 5.38 \cdot 10^{-26} \, cm^2; \; \boldsymbol{s}_a^9 = \boldsymbol{s}_a^{Pu} = 2.12 \cdot 10^{-26} \, cm^2; \tag{17}$$

$$\boldsymbol{n} = 2.9; \; \boldsymbol{t}_b \sim 3.3 \text{ days}; \; \boldsymbol{u}_n \approx 10^9 \text{ cm/s}; \; D \approx 2.8 \cdot 10^9 \text{ cm}^2/\text{s}. \tag{18}$$

The program for the solution of the system of equations (3)-(7) taking into account boundary (10)-(12), (19)-(21) and initial (13)-(15) conditions and also constants (16)-(18) was made for Fortran Power Station 4.0. At the same time DMOLCH subprogramme from the IMSL program mathematical library was used. This subprogramme DMOLCH solves a system of partial differential equations of $u_t=f(x,t,u_x,u_{xx})$ form by the method of straight lines. Results of solving of one-dimensional georeactor model are presented in Fig.1.

Obviously, the numerical solution of the system of equations (3)-(7) with different parameters confirms the fact of originating self-regulating neutron-fission wave. Although general ideology of given task solving and outputs, which take into account three-dimensional geometry, multi-group approximation for a neutron spectrum, the uranium fuel dilution and heat transmission equations, will be considered in other paper, we consider necessary to show here similar results for the simple three-dimensional model of cylindrical georeactor in order to illustrate the stability of phenomenon of self-regulating neutron-fission wave.

Without going into details of computational experiment algorithmization we note only that the net-point method in the implicit form [*Samarsky*, 1977; *Samarsky and Nikolaev*, 1978; *Samarsky and Gulin*, 2003] was used for the numerical solving of the system of partial differential equations of Eqs.(3)-(7) type describing the neutron diffusion and concentration kinetics of nuclear reaction products in cylindrical coordinates. This method does not require additional information about type of the solution of equation, and it is its main advantage.

So, let a radius $r$ and length $l$ of uranium cylinder are equal to 100 and 1000 cm, respectively, whereas all the other parameters are such as in the one-dimensional model of georeactor. Simulation data of cylindrical georeactor operation with finite length ($l =1000$ cm) and infinite radius ($r = \infty$) or, in other words, extreme case of the transition of cylindrical georeactor to one-dimensional model, are presented in Fig. 2. The total process life is 50 days. On the other hand, Fig. 3 shows similar results but for cylindrical georeactor with finite length ($l=1000$ cm) and finite radius ($r = 100$ cm). In this case to emulate neutron escape the boundary conditions are set so that gradient of the neutron concentration on the boundary of georeactor would be equal to 0.5. Physically it is equivalently to the neutron reflector with the coefficient of 0.5. Note that the iron, which is always present in the necessary quantity on the boundary of liquid and solid phases of the Earth core, can play the role of real neutron reflector. The total process life shown in Fig. 3 corresponds to 240 days. And, finally, simulation data of cylindrical georeactor operation presented in Fig.3 at the fixed time of 210 days are shown in Fig. 4. It is obvious that a spatial-temporal picture of the kinetics and distribution of the concentration of neutrons and main nuclides in radial half plane of cylindrical reactor is evidently confirms the stability of phenomenon of self-regulating neutron-fission wave. However, in any case the general proof and/or the determination of stability conditions for self-regulating neutron-fission wave in the three-dimensional medium requires the special physical and mathematical substantiation (but it is already individual problem, exceeding the limits of given work).

As nuclear energy-release is high, a considerable warming-up takes place at quite small depth of reaction. In this case the heat sink is lightened by the low velocity of neutron-fission wave and is realized by the liquid-metallic coolant (iron), which is present in the area of actinid shell on the boundary of the solid and liquid phases of Earth's core. Let us consider the nuclear-geophysical aspects of the initiation of the progressing wave of nuclear burning in real $^{238}$U-medium.

Two-phase layer $UO_2$/Fe on the surface of Earth's solid core is a natural medium for neutron-fission wave development. Since in such a wave contemporary and even depleted uranium can react, let us estimate the real possibility of wave process. The critical concentration of pure $^{239}$Pu in $^{238}$U in infinite medium, which was calculated by octa-group constants, is about 3.7 % [*Abagyan*, 1964; *Anisichkin et al.*, 2003; *Anisichkin et al.*, 2005]. Dilution by oxygen ($UO_2$/$PuO_2$) leads to the increase of critical concentration to $n_{crit} \sim 6.4$ %. The presence of iron in nuclear fuel pores (with typical "poured" concentration about 60 %) will increase the critical concentration of $^{239}$Pu up to $n_{crit} \sim 8.2$ % ($r \sim 19.5$ g/cm$^3$ for $UO_2$/$PuO_2$ and $r \sim 12$ g/cm$^3$ for Fe were used at calculations) [*Ershov and Anisichkin*, 2003]. Non-trivial thermodynamics conditions, i.e. high temperature and pressure, might rise the critical concentration of Pu up to $n_{crit}$~10 %. This means that the model system of equations (3)-(7) qualitatively closely reflects the main properties of real breeding medium, taking into account that the addition of oxygen and Fe practically does not change the solutions because their neutron-absorption cross-sections are, at least, less by the order of magnitude than the similar values for actinides.

Here the natural question arises: why $UO_2/Fe$ actinid web ($\rho$ ~15 g/cm$^3$) located on the boundary of the liquid and solid phases of the Earth's core does not sink to the center of inner core ($\rho$=12.76–13.09 g/cm$^3$ [*Anderson*, 1989]) due to gravitation instability? We believe that there are a few causes.

1. Despite the fact that the Earth's inner core was discovered 60 years ago, some seismologists, analyzing waves penetrating the inner core, still are not sure, if it is solid or liquid, or "a matter with new properties" are needed for its description [*Kuznetsov*, 1997]. They are practically convinced that an inner core (G-core according to Bullen model [*Bullen*, 1978]) is solid, but as direct proof they consider the shear-wave recording, which penetrate G-core (so called PKJKP–waves). A sole paper [*Julian et al.*, 1972] devoted to the detection of this wave was not acknowledged by seismologists. Tromp [1995] noted that PKJKP has become the Holy Grail of body-wave seismology as a figurative symbol of unsuccessful searches of this sainted bowl with Christ's blood by many generations of errant knights.

As long as PKJKP–wave existence and, consequently, the experimental value of inner core density have not the convincing proof, it is possible to suppose that model values of actinid medium (~15 g/cm$^3$) and inner core (12.76–13.09 g/cm$^3$ [*Anderson*, 1989]) are equal within the limits of 20% error.

2) Recently colossal number of seismic traces (310 000 according to the work [*Su* et al., 1994]) passing through core was analyzed and as a result the really wonderful properties of core were revealed.

It is a question of the discovery of inner core wave anisotropy, which consists in the fact that velocity of so-called PKiKP–waves have when crossing the core along the Earth's rotation axis is just a little more than velocity of the same waves have when they cross the core in the equator plane. Note that most researchers of inner core anisotropy consider that it is peculiar to relatively thin layer near core boundary [*Kuznetsov*, 1997]. Su J. and A. Dziwonski [1995] for the first time obtained the three-dimensional image of inner core anisotropy by travel-time data of 313422 traces of PKiKP–waves (registered by 2335 seismic stations from 26377 earthquakes) and showed that it amounts to few percent and is concentrated in the layer 200-300 km thick on the core boundary. Just later the Russian geophysicists [*Adushkin et al.*, 1997; Lobkovsky *at. al*, 2004] based on the information of the PKiKP –wave registration from the nuclear explosions at small epicentral distances determined that really layer thickness is much less and comes to 2–4 km. They also showed that other characteristics of inner core are no less interesting. Thus, for example, the seismic data are best explained by mosaic structure of the inner core's surface. Such a mosaic may be composed of patches, in which the transition from solid inner to liquid outer core includes a thin partially liquid layer interspersed with patches containing a sharp transition. Moreover the density of 2.2-km-thick layer [*Kracnoshchekov et al.*, 2005] corresponds to the bottom of the outer core (12.1663 g/cm$^3$) and the top of the inner core (12.7636 g/cm$^3$) for liquid and solid layers respectively, while P-wave velocity is 12 km/s [*Adushkin et al.*, 1997; *Adushkin et al.*, 2004; *Kracnoshchekov et al.*, 2005].

If this result will be confirmed by other authors, such a layer of increased density can become a platform or medium for actinid concentration (in particular, for carbides and dioxides of uranium and thorium). In this case the actinid shell as $UO_2$/Fe two-phase layer on the surface of solid (iron) core, in which iron ($r \sim 12.0$ g/cm$^3$) is in the pores of nuclear fuel ($r \sim 19.5$ g/cm$^3$) at "poured" concentration about 90%, does not sink to the center of inner core ($r$=12.76–13.09 g/cm$^3$ [*Anderson*, 1989]) due to gravitation instability as it has density ~12.75 g/cm$^3$. This, in its turn, leads to the increase of critical concentration to $n_{crit} \sim 10$-12 %. It is obvious, that such change of two-phase layer density and critical concentration, respectively, practically in no way will not change previous results on Feoktistov's neutron-fission wave simulation.

The question of not less importance is: "Where do neutrons come from for chain reaction initiation?" In spite of the active discussions of the possibility of chain nuclear reaction existence in interior of the Earth and other planets in numerous papers (starting with Kuroda [1956] and ending with Driscoll [1988], Herndon [1993, 1996], Anisichkin et al. [1997, 2003, 2005], the question of the natural external neutron sources, which locally start the mechanism of nuclear burning, remains open and requires serious joint efforts of the theorists.

However, taking into account all difficulties concerning the explanation of the mechanism of neutron-fission wave starting, it is possible to take an alternative route and to try to find in the thermal history of the Earth geophysical events, which directly or indirectly denote the existence of slow nuclear burning. Note that these events should be in recent times, which as the present, characterized by lowered, i.e. subcritical concentration of odd isotopes of uranium and plutonium. Let us consider below the example of such geophysical paleoevents.

### 3. $^3$He/$^4$He-ratio distribution in the Earth's interior as quantitative criterion of georeactor thermal power

Fundamental models of anomalous $^3$He concentration origin and $^3$He/$^4$He-ratio distribution in the Earth's interior have serious contradictions. Without going into details, we cite Anderson [1998] who, in our opinion, reproduces closely the current state of problem: "The model whereby high $^3$He/$^4$He is attributed to lower-mantle source, and is thus effectively an indicator of plumes, is becoming increasingly untenable as evidence for a shallow origin for many high-$^3$He/$^4$He hotspots accumulates. Shallow, low-$^4$He for high $^3$He/$^4$He are logically reasonable, cannot be ruled out, and need to be rigorously tested if we are to be understand the full implications of this important geochemical tracer".

In our case we suppose that $^3$He is produced by natural reactor located on the boundary of liquid and solid phases of the Earth's core. At the same time $^4$He is produced both by georeactor and due to decay of $^{238}$U and $^{232}$Th in the crust, the upper (depleted) and lower mantle of the Earth.

To determine $^4$H?-accumulation rate we used total and partial radiogenic heat production rates of uranium $H_U$ and thorium $H_{Th}$ in the crust, the upper (depleted) mantle and directly in the mantle (Table 1), which have been earlier received [*Rusov et al.*, 2003] within the framework of O'Nions-Evensen-Hamilton geochemical model [*O'Nions et al.*, 1979]. Note that these estimations are very close to estimations obtained within the framework of the well known Bulk Silicate Earth model [*Hofmeister and Criss,* 2005] (see, for example, the works of Fiorentini et al. [2004, 2005]). At the same time our model [*Rusov et al.*, 2003] as well as others "models of Earth's bulk composition based on CI chondritic meteorites provide an unrealistically low radioactive power of ~ 20 TW" [*Hofmeister and Criss*, 2005] in comparison with heat flow observed now (frequently quoted estimate is $H_E$= (44±1) TW [*Pollack et al.*, 1993]).

It is obvious that a difference between a real heat (which is produced now in the Earth) and a calculated heat (i.e. a radiogenic heat in frameworks the Bulk Silicate Earth model [*Hofmeister and Criss*, 2005]) can be very significant even with allowance for the high thermal inertia of the Earth ($t_E \approx 10^9$ years [*Van den Berg and Yuen* (2002); *Van den Berg et al.*, 2002]). Anderson [2005] refers to this difference as the missing heat source problem and summarizes the situation with following words: "Global heat flow estimates range from 30 to 44 TW… Estimates of the radiogenic contribution (from the decay of U, Th and K in the mantle), based on cosmochemical considerations, vary from 19 to 31 TW. Thus, there is either a good balance between current input and output… or there is a serious missing heat source problem, up to a deficit of 25 TW…"

In any case the decisive argument in favour of one or another paradigm can be only experiment (trivial as it may seem). Since radiogenic component is essentially based on cosmochemical considerations, which, as it is well known, cause uncertainty, only a direct determination as offered by geo-neutrino detection or the indirect determination of $^3$He/$^4$He-ratio depth distribution are important. In other words, if one can determine the amount of radioactive elements by means of geo-neutrinos and/or $^3$He/$^4$He-ratio, an important ingredient of the Earth's energetics will be fixed [*Fiorentini et al.*, 2004; *Fiorentini et al.*, 2005].

We consider that such an additional thermal power (designated as $H_f$ ) can be caused by nuclear burning of the actinid shell, which consists of chemically stable and high density actinid compounds. As it is shown experimentally in the works of Anisichkin et al. [2003, 2005], these compounds could be lowered together with melted iron and concentrate on the surface of inner solid core ($r_N \approx 1200$ km) due to substance gravity differentiation.

It is obvious, that if this thermal power $H_f$ is generated only owing to radiogenic heat $H_a$, there will be no contribution of actinide shell to the geoantineutrino integral intensity. In order to obtain the real contribution of actinide shell we suppose that the energy-release power $H_f$ of actiniod shell as a nuclear energy source is essentially higher than the partial power of radiogenic heat $H_a$, produced by $^{238}$U and $^{232}$Th radioactive chains, i.e. $H_a << H_f$.

For simplicity sake, further we consider the actinid shell as $UO_2$/Fe two-phase layer on the surface of solid (iron) core of the Earth. Iron ($r \sim 12.0$ g/cm$^3$) in the pores of nuclear fuel ($r \sim 19.5$ g/cm$^3$), whose "poured" concentration is $\sim 90\%$, decreases the two-phase layer density to $\sim 12.75$ g/cm$^3$. Let us assume $H_a \sim 0.1 \div 0.5$ TW. If the two-phase actinid medium with the total mass of natural uranium

$$M(U) = \frac{H_a}{e(U)} \sim 10^{15} kg, \quad where \quad e(U) \cong 0.95 \cdot 10^{-4} \, W/kg, \qquad (19)$$

represents a continuous homogeneous shell on the surface of the Earth's solid core, its thickness will be $\sim 1 \div 5$ cm. Apparently, it is more correct to image such a two-phase actinid medium as the inhomogeneous shell, which represents the stochastic web of actinid "rivers" and "lakes" located in the valleys of rough surface [Anderson, 1989; Lobkovsky et al., 2004] of the Earth's solid core.

Bellow we consider georeactor model of the origin of $^3$He anomalous concentrations and the $^3$He/$^4$He-ratio distribution in the Earth's interior. If the existence of a georeactor will be experimentally confirmed, this model naturally explains the so-called helium paradoxes [Anderson, 1998].

So, let us assume that a reactor power is equal to $P=30$ TW. The further calculations in the framework of georeactor model will show that this value is the most adequate valuation of reactor power. In our case, the marvelous constancy of anomalous isotopic composition of the mantle helium is explained by the properties of fast ($\sim 1$ MeV) neutron-induced fission of $^{239}$Pu in neutron-fission wave front. The $^3$H? production probability is mainly determined by the probability of $^3$H production as fission fragment of $^{239}$Pu triple fission. This probability is about $\sim 1.6 \cdot 10^{-4}$ [Vorob'ev et al., 1974]. Hence, the total accumulation rate of $^3$H? produced due to tritium $b$−decay (T$_{1/2}$ $\sim 12.3$ years) is approximately equal:

$$N_{fB}^{30}(^3He) \sim 1.6 \cdot 10^{-4} \, n_f \approx 1.6 \cdot 10^{-4} \frac{P}{E_f} = 14.8 \cdot 10^{19} \, s^{-1}, \qquad (20)$$

where $E_f = 210.3$ MeV is the average energy per $^{239}$Pu fission.

On the other hand, $^4$H? accumulation rate due to $^{238}$U radioactive decay in UO$_2$/Fe actinide web (by hypothesis of $H_a^U \approx 0.1 \div 0.5$ TW) has the form:

$$N_{fB}(^4He) \sim 8 \frac{H_a^U}{Q_a^U} = 8 \frac{(0.1 \div 0.5) \cdot 10^{12}}{51.7 \cdot 1.6 \cdot 10^{-13}} \approx (9.7 \div 48.5) \cdot 10^{22} \, s^{-1}. \qquad (21)$$

So, helium ratio $R_{fB}$ in UO$_2$/Fe actinide web (located on the boundary of solid and liquid phase of the Earth's core) is equal:

$$R_{fB}^{30} = \frac{N_{fB}^{30}(^3He)}{N_{fB}(^4He)} \approx (0.3 \div 1.6) \cdot 10^{-3}. \qquad (22)$$

Here and in future, we use a number of physical suppositions, which make it possible (without loss of generality) to obtain the rough estimations of helium ratio $R$ for different geospheres of the Earth. At first, the simplified consideration of helium isotopes transport process in the medium is connected with supposition that a radial drift dominates over diffusion and average radial drift speeds of $^3$H? and $^4$H? in are approximately equal in gravity field of different geospheres of the Earth. At the same time the average cross-sections (or probabilities) of these isotopes capture by different traps (bags, bed joints, rock pores etc.) in the Earth are also approximately equal, but they are so small, that we can neglect the decrease of these isotopes flows in the direction of radial drift.

Now we can estimate the $R$ ratio in the mantle and crust. Earlier [*Rusov et al.*, 2003] on basis O'Nions-?vensen-Hamilton geochemical model [*O'Nions et al.*, 1979] the integral estimates of thermal flux from uranium ($H_U$=5.1 TW) and thorium ($H_{Th}$=5.7 TW) in the mantle was obtained (see Table 1). Then $^4$H? accumulation rate due to $^{238}$U and $^{232}$Th radioactive decay in the mantle (minus the depleted upper mantle ) will be approximately equal:

$$N_{M-UM}(^4He) \sim 8 \frac{H_U}{Q_a^U} + 6 \frac{H_{Th}}{Q_a^{Th}} \approx 9.93 \cdot 10^{24} \, s^{-1}, \qquad (23)$$

where $Q_a^U$ =51.7 MeV and $Q_a^{Th}$ =42.8 MeV are decay energies:

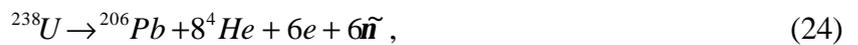

$$^{238}U \rightarrow {}^{206}Pb + 8\,^4He + 6e + 6\tilde{n}, \qquad (24)$$

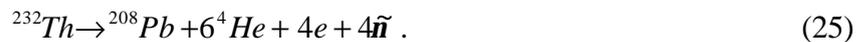

$$^{232}Th \rightarrow {}^{208}Pb + 6\,^4He + 4e + 4\tilde{n}. \qquad (25)$$

Therefore, helium ratio in the mantle (minus the depleted upper mantle) $R_{M-DUM}$ due to Feoktistov reactor operating ($P$= 30 TW) with allowance for Eqs. (20), (21) and (23) is approximately equal

$$R_{M-DUM}^{30} = \frac{N_{fB}^{30}(^3He)}{N_{M-DUM}(^4He) + N_{fB}(^4He)} \cong \frac{N_{fB}^{30}(^3He)}{N_{M-DUM}(^4He)} \approx 11.15 R_a, \qquad (26)$$

where $R_a$=1.38·10⁻⁶ is atmospheric helium ratio.

In a similar manner, the average values of helium ratio for upper (depleted) mantle $R_{UM}$ and for crust $R_{crust}$ look like

$$R_{DUM}^{30} = \frac{N_{fB}^{30}(^3He)}{N_{DUM}(^4He) + N_{M-DUM}(^4He) + N_{fB}(^4He)} \approx 9.05 R_a, \qquad (27)$$

$$R_{crust}^{30} = \frac{N_{fB}^{30}(^3He)}{N_{crust}(^4He) + N_{DUM}(^4He) + N_{M-DUM}(^4He) + N_{fB}(^4He)} \approx 7.55 R_a, \qquad (28)$$

At the same time the statistical analysis of the deep distribution of helium isotopes made on the basis of numerous experimental data has shown that the average value of helium ratio for the crust $R_{crust}$ and upper mantle $R_{DUM}$ are $R_{crust}$=(7.47±1.95)$R_a$ and $R_{DUM}$=(9.14±3.59)$R_a$, whereas values $R_{M-UM}$=(11÷15)$R_a$ are commonly attributed to deep mantle plumes and "indicative of lower mantle involvement" [*Anderson*, 2000]. It is obvious, that theoretical estimates (26)-(28) practically coincide with experimental data.

Thus, if Feoktistov reactor power is 30 TW, the average values of helium ratios for crust, upper mantle, mantle (minus the depleted upper mantle) and a thin layer on the boundary of liquid and solid Earth's core $R_{fB}$ come to following values:

$$R_{crust}^{30} \approx 7.6 R_a, \quad R_{DUM}^{30} \approx 9.1 R_a, \quad R_{M-DUM}^{30} \approx 11.2 R_a, \quad R_{fB}^{30} \approx (220 \div 1160) R_a, \qquad (29)$$

which are in close agreement with the corresponding average values of experimental helium ratios [*Anderson*, 2000] too.

At last, considering some lower layer of undepleted mantle ($M-DUM$) as the area of the lower mantle ($LM$), whose characteristic volume is $V_{LM} \cong (0.2-0.3) V_{M-DUM}$ (where so-called Morgan's plumes [*Morgan*, 1971] probably originate) it is possible within the framework of our model to obtain the average value of helium ratios $R_{LM}$ for the lower mantle:

$$R_{LM}^{30} = \frac{V_{M-DUM}}{V_{LM}} R_{M-DUM}^{30} = (30 \div 50) R_a, \qquad (30)$$

which agree to a high accuracy with well known experimental data by Stuart et al. [2003] and Kellogg and Wassenburg [1990].

Thus, we have obtained a very important result, because if $^3$H? in reality has a georeactor origin, the $^3$He/$^4$He ratio distribution in the Earth's interior is natural quantitative criterion of georeactor thermal power. Moreover, if the georeactor exists, the corresponding $^3$He/$^4$He ratio distribution is predetermined not only by georeactor thermal power but also by the corresponding distribution of $^{238}$U and $^{232}$Th in the crust and mantle, which is correctly calculated in the work of Rusov et al. [2003].

## 4. Contribution of georeactor antineutrinos
## to the antineutrino spectrum of the Earth. Comparison with the experiment.

It is obvious that the unambiguous test for georeactor existence in the Earth's interior is the geoneutrino spectrum (especially at energies >3.272 ? eV, where only "fission geoneutrinos" are detected, i.e., geoneutrinos produced due to actinid fission). In this sense the georeactor idea is fruitful not only for understanding of true physical essence of so-called "helium paradoxes" [*Anderson*, 1998, 2000], but at the same time it effectively solves the description problem of geoantineutrino spectrum and reactor antineutrino experimental spectrum in KamLAND in the range of antineutrino energy ~ 2.8 MeV (see Fig. 3 in paper of *Araki et al.,* 2005).

So, the $^{239}$Pu fission rate in neutron-fission wave front is

$$\boldsymbol{h}_f = P \big/ E_f \approx 8.9 \cdot 10^{-23} \ \text{fission/s}, \tag{31}$$

where $E_f = 210.3$ MeV is the average energy per $^{239}$Pu fission.

Hence, the crude estimation of antineutrino integral intensity in two diametrically opposite points on the Earth's surface from burning wave front in $UO_2$/Fe actinides web has the form:

$$\Phi_{\tilde{n}} = \frac{1}{4\boldsymbol{p}(R_\oplus \pm r_N)^2} \cdot \boldsymbol{h}_f \cdot \boldsymbol{m}_{\tilde{n}} = \begin{cases} 0.72 \cdot 10^6 \ cm^{-2} s^{-1}, & if \quad "+", \\ 1.56 \cdot 10^6 \ cm^{-2} s^{-1}, & if \quad "-". \end{cases} \tag{32}$$

where $\boldsymbol{m}_{\tilde{n}} \approx 5.7$ is the number of antineutrinos per $^{239}$Pu fission; $R_\oplus \approx 6400$ km; $r_N \approx 1200$ km.

Using the design procedure of partial and total energy $\boldsymbol{b}$-, $\tilde{\boldsymbol{n}}$-spectra of radioactive nuclides [*Rusov et al.*, 2003] we have constructed the partial $d\Phi_{\tilde{n}}/dE$ ($^{238}$U), $d\Phi_{\tilde{n}}/dE$ ($^{232}$Th), $d\Phi_{\tilde{n}}/dE$ ($^{40}$K) (Fig.5) [*Rusov et al.*, 2003], $d\Phi_{\tilde{n}}/dE$ ($^{239}$Pu) (Fig.6) [*Rusov et al.*, 2004?] and the antineutrino total energy spectra (without oscillations) of the Earth $d\Phi_{\tilde{n}}/dE$ ($^{238}$U+$^{232}$Th+$^{40}$K+$^{239}$Pu) [*Rusov et* *al.*, 2004?] (Fig. 7). The partial contributions were previously normalized to corresponding geoantineutrino integral intensity on the Earth's surface [*Rusov et al.*, 2003, 2004?]).

The theoretical form of measured total energy spectrum $dn_{\tilde{n}}/dE$ (Fig.4) looks like

$$\frac{dn_{\tilde{n}}}{dE} = \boldsymbol{e} N_p \sum_i p_i (E_{\tilde{n}}, L) \frac{d\boldsymbol{l}_{\tilde{n}}^i}{dE} \boldsymbol{s}_{np} (E_{\tilde{n}}) \Delta t, \qquad \text{MeV}^{-1}, \qquad (33)$$

where probability of neutrino oscillation can be written for two neutrino flavours as

$$p(E_{\tilde{n}}, L) \cong 1 - \sin^2 2\boldsymbol{q}_{12} \sin^2 \left( \frac{1.27 \Delta m_{12}^2 [eV^2] L[m]}{E_{\tilde{n}} [MeV]} \right), \qquad (34)$$

here $d\boldsymbol{l}_{\tilde{n}}/dE \equiv d\Phi_{\tilde{n}}/dE$ at $E_{\tilde{n}} \geq 1.804$ MeV, $\boldsymbol{s}_{np}$ is antineutrino-proton interaction cross-section of inverse $\boldsymbol{b}$-decay reaction with corresponding radiation corrections [*Vogel*, 1984; *Fayans*,1985; *Vogel and Beacom*, 1999]; $L$ is the distance apart source and detector; $\Delta m_{12}^2 \equiv \left| m_2^2 - m_1^2 \right|$ is the mass squared difference, $\boldsymbol{q}$ is mixing angle.

In the same time, since reactor geoantineutrinos are in the spectral region of prompt energy above 2.6 MeV, the calculations of true antineutrino spectrum and oscillation parameters ($\Delta m_{12}^2$, $\sin^2 2\boldsymbol{q}_{12}$) by the traditional way in KamLAND-experiment need supplement to a definition. In other words, the traditional method of obtaining consistent estimates, for example, maximum-likelihood method, usually used for determination of oscillation parameters ($\Delta m_{12}^2$, $\sin^2 2\boldsymbol{q}_{12}$) must take into account one more reactor in the experiment or, more specifically, take into account the antineutrino spectrum of georeactor with the power of 30 TW, which is located at a distance of $L \sim 5.2 \cdot 10^6$ m. The results of such approach will be described in our next paper, whereas we offer here the simple estimation approach. The results of its application show that hypothesis of existence of the georeactor with the power of 30 TW on the boundary of liquid and solid phases of the Earth's core does not conflict with the experimental data.

We used the next circumstance. If CPT invariance is assumed, the probabilities of the $\boldsymbol{n}_e \rightarrow \boldsymbol{n}_e$ and $\tilde{\boldsymbol{n}}_e \rightarrow \tilde{\boldsymbol{n}}_e$ oscillations should be equal at the same values $L/E_{\boldsymbol{n}}$. At the average distance $L \sim 180$ km of the Japan reactors from the KamLAND detector and the typical energies of a few MeV of the reactor $\tilde{\boldsymbol{n}}_e$, the experiment has near optimal sensitivity to the $\Delta m^2$ value of the LMA solar solution [*Barger et al.*, 2003]. Now $\boldsymbol{i}$ is known that the mass squared difference indicated by the solar neutrino data is $\sim 6 \cdot 10^{-5}$ eV$^2$ and the mixing is large but not maximal, $\tan^2 \boldsymbol{q} \sim 0.4$ [*Achmed et. al.*, 2004].

Because the sensitivity in $\Delta m^2$ can dominate by the spectral distortion in the antineutrino spectrum, while solar neutrino data provide the best constraint on $\boldsymbol{q}$, within the framework of further analysis we can suppose (basing on CPT-theorem) that the angle of mixing in KamLAND - experiment is determined by the "solar" equality $\tan^2 \boldsymbol{q}_{12} = 0.4$ or $\sin^2 2\boldsymbol{q}_{12} = 0.83$. Therefore to calculate the integral intensity of reactor geoneutrinos the following approximation for survival probability $p_{i=\mathrm{Pu}}$ of Eq. (34) type was also used:

$$p_{i=Pu}(E_{\tilde{n}}, L) = 1 - 0.5 \sin^2 2\boldsymbol{q}_{12} \cong 0.59, \quad L \gg L_{osc}[m] = \frac{2.48 E_{\tilde{n}}[MeV]}{\Delta m_{12}^2 [eV^2]}, \quad (35)$$

where $L_{osc}$ is the oscillation length, $L \sim 5.2 \cdot 10^6$ m is the distance apart the boundary of liquid and solid phases of the Earth's core and detector.

Then using Eq.(33) it is possible to show that in first KamLAND-[*Eguchi et al.*, 2002; *Eguchi et al.,* 2003] the integral intensity of reactor geoantineutrinos $^{Pu}n_{\tilde{n}}$ on the Earth's surface with consideration of Eq. (35) is equal

$$^{Pu}n_{\tilde{n}} = p_{i=Pu} \cdot \boldsymbol{e} \cdot N_p \cdot \Delta t \cdot \int\limits_{E=1.804}^{\infty} \frac{d\boldsymbol{I}_{\tilde{n}}(Pu)}{dE} \cdot \boldsymbol{s}_{np}(E) dE =$$

$$= p_{i=Pu} \left[ 12.72(Pu)\big|_{E_n \le 3.272} + 30.24(Pu)\big|_{E_n > 3.272} \right] = 7.50\big|_{E_{\tilde{n}} \le 3.272} + 17.84\big|_{E_{\tilde{n}} > 3.272}, \quad (36)$$

where $\boldsymbol{e} \approx 0.783$ is detection efficiency; $N_P = 3.46 \cdot 10^{31}$ is the number of protons in the detector sensitive volume; $\Delta t = 1.25 \cdot 10^7$ s is exposure time [*Eguchi et al.*, 2002; *Eguchi et al.,* 2003]; $\boldsymbol{s}_{np}$ is antineutrino-proton interaction cross-section of inverse $\boldsymbol{b}$-decay reaction with corresponding radiation corrections [*Vogel,* 1984; *Fayans*,1985; *Vogel and Beacom,* 1999].

Now for the domain of the prompt energies $E_{prompt} > 2.6$ MeV (see Fig. 8a) we determine the ratio of "true" flux of reactor antineutrinos $N_{obs}$, which is equal to difference of the measured flux $N_{full}$ and background caused by $^{13}$C($\boldsymbol{a}$, $n$)$^{16}$O reaction [*Araki et al.*, 2005], $N_C$ and reactor geoneutrinos $^{Pu}n_{\tilde{n}}$ to expected flux $N_{expected}$ in KamLAND-experiment. Taking into account that in first KamLAND-experiment $N_{full} = 54$, $N_{expected} = 86.8 \pm 5.6$ [*Eguchi et al.*, 2002; *Eguchi et al.,* 2003], $N_C \cong 2$ (see Fig. 8?), $^{Pu}n_{\tilde{n}}$ ($E_{prompt} > 2.6$ MeV)=17.84 (see Eq. (36)), the ratio $\Re$ is equal

$$\Re = \frac{N_{obs}}{N_{expected}} = \frac{N_{full} - N_C - {}^{Pu}n_{\tilde{n}}}{N_{expected}}\Bigg|_{E_{prompt} > 2.6} \cong 0.394 \pm 0.096(stat) \pm 0.042(syst). \quad (37)$$

The probability that the KamLAND result is consistent with the no disappearance hypothesis is less than 0.05%. Fig.9 shows the ratio $\Re$ for KamLAND as well as previous reactor experiments as function of the average distance from source. Ibidem the shaded region, which indicates the range of flux predictions corresponding to the 95% C.L, is shown. LMA region found in a global analysis of the solar neutrino data [*Fogli et al.*, 2002]. It appears that only those values, which are in interval $\Delta m_{12}^2 \approx (2\div4)\cdot10^{-5}$ eV$^2$ (Fig.10), are permitted for the given value of $\Re$ (37) and fixed angle of mixing ($\sin^2 2q_{12}$=0.83). We chose the value of $\Delta m_{12}^2 =2.5\cdot10^5$ eV$^2$ for the further calculations. The corresponding shape of antineutrino spectrum at given $\Re$ (37) (see the insert in Fig.10), which was calculated for first KamLAND-experiment at the fixed angle of mixing and different $\Delta m_{12}^2$ from the interval $(2\div4)\cdot10^{-5}$ eV$^2$, was used as a rule of selection of this value.

Calculations of theoretical antineutrino spectrums at the given oscillation parameters (see the insert in Fig.10 and Fig. 8b) was made by Eqs.(33)-(34). Necessary parameters characterizing exposure time, detection geometry and detector properties are taken from KamLAND-experiment data [*Eguchi et al.*, 2002; *Eguchi et al.,* 2003]. To determine the averaged fission number of the four main nuclei ($^{235}U$, $^{238}U$, $^{239}Pu$, $^{241}Pu$) inducing the antineutrino contributions from fission products of each Japanese reactors in the radius 1000 km from detector we took the parameters necessary for computation, for example, the relative fission yields ($^{235}U : {}^{238}U : {}^{239}Pu : {}^{241}Pu$) and also distances to KamLAND-detector for each of indicated groups of reactors, from Internet-site [*KamLAND Collaboration*, 2005].

Obviously, that approximate values of oscillation parameters ($\sin^2 2q_{12}$=0.83, $\Delta m_{12}^2 =2.5\cdot10^5$ eV$^2$) obtained in this way make it possible by the similar calculation procedure to determine the total geoneutrino spectrum (Fig. 8?), which includes events due to $a$-decay of $^{238}$U and $^{232}$Th (with the known radial profile of their distribution in the Earth's interior [*Rusov et al.*, 2003]) and $^{239}$Pu fission in the georeactor core, and to determine the geoneutrino integral intensity on the Earth's surface, respectively:

$$n_{\tilde{n}} = p_i \cdot e \cdot N_p \cdot \Delta t \cdot \int\limits_{E=1.804}^{\infty} \frac{dI_{\tilde{n}}(U+Th+Pu)}{dE} \cdot s_{np}(E)dE =$$

$$= p_i \left[2.70(U) + 0.78(Th) + 12.72(Pu)\big|_{E_n \leq 3.272} + 30.24(Pu)\big|_{E_n > 3.272}\right] =$$

$$= \left[2.10(U+Th) + 7.50(Pu)\big|_{E \leq 3.272} + 17.84\big|_{E>3.272}\right]. \qquad (38)$$

The found total geoneutrino spectrum (Fig. 8a, green shaded region), in its turn, makes it possible to determine the "true" antineutrino spectrum (Fig. 8b, blue points with bars) detected from the Japanese reactors in geometry of first KamLAND -experiment [*Eguchi et al.*, 2002; *Eguchi et al.,* 2003]. In Fig. 8b is also shown approximate fit oscillation, i.e. the theoretical antineutrino KamLAND-spectrum with the approximate oscillation parameters $\sin^2 2q_{12}$=0.83 ? =2.5·$10^5$ $eV^2$. Note that some difference of expected no oscillation spectrum shown in Fig. 8 from similar KamLAND-spectrum [*Eguchi et al.*, 2002; *Eguchi et al.,* 2003] is explained, apparently, by non-identity of used databases and does not exceed 3% (see Fig. 6). For computation of antineutrino spectrums of actinides we used the ORIGEN-S module of the SCALE-4.4 package 4 [*Ryman, Hermann*, 2000] and also ENSDF [*Tuli*, 2001] and ENDF-349 [*England and Rider*, 1993] nuclear data libraries.

In conclusion we give the results of oscillation parameters verification within the framework of test problem of comparison of theoretical (which takes into account the georeactor operation) and experimental spectrum of reactor antineutrino on the base of new data [Araki et al.] obtained by experimental investigation of geologically produced antineutrinos with KamLAND. For example, the new KamLAND–data [*Araki et al.,* 2005] handling in energy range $E_n$=(1.7–3.4) MeV (exposure time $\Delta t$= (749.1 ± 0.5) days, detection efficiency $e$ ≈0.687 and the number of protons in detector sensitive volume $N_P$=(3.46±0.17)·$10^{31}$) shows that obtained antineutrino spectrum, which takes into account georeactor antineutrinos, and predicted KamLAND-spectrum are practically similar (Fig. 11). In Fig.12 the theoretical (which takes into account the georeactor operation) reactor antineutrino spectrum calculated on the base of new data [*Araki et al.*, 2005] for all energy range of event detection is presented.

In conclusion it is necessary to note that although hypothesis of nuclear georeactor existence, which we used for interpretation of KamLAND-experiment, seems to be very effective, it can be considered only as a possible alternative variant of KamLAND experimental data description. Only direct measurements of geoantineutrino spectrum in the energy range >3.4 ? eV in future underground or submarine experiments will finally solve the problem of natural georeactor existence and will make it possible to determine the "true" values of reactor antineutrinos oscillation parameters.

## 5. Conclusions

Based on the analysis of the temporal evolution of radiogenic heat-evolution power of the Earth within the framework of the geochemical model of the mantle differentiation and the Earth's crust growth [*Rusov et. al.*, 2003; *O'Nions et al.*, 1979] supplied by a nuclear energy source on the boundary of the solid and liquid phases of the Earth's core, we have obtained the tentative estimation of geoantineutrino intensity and geoantineutrino spectrum on the Earth surface from different radioactive sources ($^{238}$U, $^{232}$Th, $^{40}$K and $^{239}$Pu).

We have also showed that natural nuclear reactors may exist on the boundary of the solid and liquid phases of the Earth's core as spontaneous reactor-like processes with U−Pu (Feoktistov's fuel cycle) and/or Th−U (Teller-Ishikawa-Wood's fuel cycle). Note that, as compared to $^{238}$U-medium, the wave velocity in $^{232}$Th-medium has the value about $L/t \sim 0.1$ cm/day (where L~5 cm is the diffusion length of neutron absorption thorium, $t$ =39.5/ln2≈57 days is time of $^{233}$U generation due to $b$−decay of $^{233}$Pa). It means that speed of neutron-fission wave propagation in $^{232}$Th-medium (Teller-Ishikawa-Wood's fuel cycle) is less by an order of magnitude than the similar speed of Feoktistov's burning wave.

The solution of the main problem connected with the search of natural neutron sources, which locally start the mechanism of nuclear burning, is unclear and (in spite of the active discussions of the possibility of the existence of chain nuclear reaction in interior of the Earth and other planets in the numerous papers) requires a serious joint efforts of the theorists.

However, taking into account all difficulties concerning the explanation of the mechanism of neutron-fission wave starting, it is possible to go by an alternative route and to try to find in the thermal, seismic or magnetic history of the Earth such geophysical events, which directly or indirectly denote the existence of slow nuclear burning. First of all it concerns, apparently, such geophysical phenomena as anomalous $^3$H/$^4$H-ratio distribution in the Earth's interior and a geoneutrino spectrum on daylight of the Earth (KamLAND-experiment). It turned out, that in both cases the presence of a georeactor (as nuclear burning progressing wave) makes it possible to obtain the model $^3$H/$^4$H-ratio distribution and a geoneutrino spectrum, which are in good agreement with experimental data.

At last, it is necessary to note that Feoktistov's burning wave can effectively provide the convective mechanism of the sustained Earth's hydromagnetic dynamo operation, as it naturally creates conditions for gravity convection in the liquid core caused by the effective floating up of light fission fragments behind the nuclear burning wave front. It is an important point, as the condition of continual sustained weak (when temperature is close to adiabatic) convection in liquid core is the cause and condition of differential rotation of the different layers of core and, consequently, the geomagnetic field.

Thus the hypothesis of slow nuclear burning on the boundary of the liquid and solid phases of the Earth's core is very effective for the explanation of some features of geophysical events. However, strong evidences can be obtained from the independent experiment on geoantineutrino energy spectrum measurements using the multi-detector scheme of geoantineutrino detection on large base. At the same time the solutions of the direct and inverse problems of neutrino remote diagnostics of the intra-terrestrial processes connected with the obtaining of pure geoantineutrino spectrum [*Rusov et al*., 2004b] and the correct determination of $b$-sources radial profile in the Earth's interior will undoubtedly help to solve the problems both of the existence of natural nuclear reactor on the boundary of the liquid and solid phases of the Earth's core and true geoantineutrino spectrum.

Table 1. Mass distribution, antineutrino fluxes and heat production rates

($M$, $\boldsymbol{F}$ and $H$ are in units of $10^{17}$ kg, $10^6$ cm$^{-2}$s$^{-1}$ and TW, respectively)

[*Rusov et al.*, 2003]

| Geospheres | $^{238}$U | | | $^{232}$Th | | | $^{40}$K | | | $H$ |
|---|---|---|---|---|---|---|---|---|---|---|
| | $^iM$ | $\boldsymbol{F_{\bar{n}}}$ | $H$ | $^iM$ | $\boldsymbol{F_{\bar{n}}}$ | $H$ | $^iM$ | $\boldsymbol{F_{\bar{n}}}$ | $H$ | |
| Crust | 0.22 | 1.040 | 2.10 | 0.55 | 0.57 | 1.50 | 0.271 | 4.60 | 0.97 | 4.6 |
| Depleted mantle | 0.06 | 0.170 | 0.60 | 0.59 | 0.36 | 1.60 | 0.0094 | 0.95 | 0.04 | 2.2 |
| Mantle | 0.53 | 0.992 | 5.10 | 2.10 | 0.87 | 5.70 | 0.53 | 3.57 | 1.90 | 12.7 |
| $\Sigma$ | 0.81 | 2.20 | 7.80 | 3.24 | 1.80 | 8.80 | 0.81 | 9.12 | 2.90 | 19.5 |

# Figure captions

**Fig.1.** Concentration kinetics of (a) neutrons; (b) $^{238}$U; (c) $^{239}$U; (d) $^{239}$Pu in the active zone of one-dimensional georeactor. Here $t$-line is time axis, step is $\Delta t$=0,01 s; $x$-line is spatial coordinate axis, step is $\Delta x$=1 cm; $z$-line is concentration axis, particle/cm$^3$, $\boldsymbol{F_0}$=5·10$^{17}$cm$^{-2}$s$^{-1}$. As the boundary condition Eq. (12) was used.

**Fig.2.** Concentration kinetics of (a) neutrons; (b) $^{238}$U; (c) $^{239}$U; (d) $^{239}$Pu in the active zone of cylindrical georeactor with infinite radius (one-dimensional case emulation) and 1000 cm long. Here $t$-line is time axis, step is $\Delta t$ = 0,1 day; $x$-line is spatial coordinate axis, step is $\Delta x$ =1 cm; $z$-line is concentration axis, particle/cm$^3$. Computation was performed with time step of 100 seconds, but only every hundredth step was recorded (500 steps is recorded in all, that equal to the 50 days). The coordinate $t$=0 corresponds to the power up time of external source. $\boldsymbol{F_0}$=1.5·10$^{19}$cm$^{-2}$s$^{-1}$.

**Fig.3.** Concentration kinetics of (a) neutrons; (b) $^{238}$U; (c) $^{239}$U; (d) $^{239}$Pu on the axis of cylindrical reactor ($r$=0) with radius of 100 cm and 1000 cm long. Here $t$-line is time axis, step is $\Delta t$ = 0,2 day; $z$-line is longitudinal spatial coordinate axis, step is $\Delta z$ =1 cm; $Z$-line is concentration axis, particle/cm$^3$.Computation was performed with time step of 100 seconds, but only every two hundredth step was recorded (700 steps is recorded in all, that equal to the 240 days). The coordinate $t$=0 corresponds to 100 days from the power up time of external source in cylinder butt-end. $\boldsymbol{F_0}$=3·10$^{17}$cm$^{-2}$s$^{-1}$.

**Fig.4.** Concentration distribution of (a) neutrons; (b) $^{238}$U; (c) $^{239}$U; (d) $^{239}$Pu in the cylindrical reactor shown in Fig. 3 but at fixed time of 210 days. Here $r$–line is transverse spatial coordinate axis, step is $\Delta r$=1 cm; $z$-line is longitudinal spatial coordinate axis, step is $\Delta z$ =1 cm; $Z$-line is concentration axis, particle/cm$^3$.

**Fig.5.** The expected $^{238}U$, $^{232}Th$ and $^{40}K$ decay chain electron antineutrino energy distributions. KamLAND-detector can only detect electron antineutrinos to the right of the vertical dotted line.

**Fig.6.** Calculated partial antineutrino spectra of $^{239}$Pu normalized to nuclear decay (a) and its deviation from theoretical spectra obtained by different authors in the energy range = 1.8 -10.0 MeV (b).

**Fig. 7.** Calculated total geoantineutrino spectrum of the Earth (taking into account the reactor power of 30 TW) in KamLAND detector. Solid line is ideal spectrum, histogram is spectrum with the energy bin of 0.425 MeV (a) and 0.17 MeV (b). Insert, the same spectrum, but for reactor power of 2.5 TW.

**Fig.8.** (a): energy spectrum of the observed prompt events (solid black circles with error bars) [*Eguchi et al.*, 2003], along with the expected no oscillation spectrum (histogram, with events from $^{13}$C($\alpha$,n)$^{16}$O reactions and accidentals shown) and calculated total geoantineutrino oscillation spectrum in KamLAND detector (green histogram); (b): ?nergy spectrum of the observed prompt neutrinos (solid blue circles with error bars), which is equal to difference between the ?nergy spectrum of the observed prompt events (solid black circles with error bars), background and total geoantineutrino (oscillation) spectrum (green histogram). Fit oscillation (lower yellow histogram) describing the expected oscillation spectrum from Japan's reactor. Vertical dashed line corresponds to the analysis threshold at 2.6 MeV.

**Fig. 9.** The ratio ($\Re = N_{\text{obs}}/N_{\text{expected}}$) of measured to expected flux from reactor experiments [*Particle Data Group*, 2002]. The solid dot [*Eguchi et al.*, 2002; *Eguchi et al.*, 2003] and circle is the KamLAND point plotted at a flux-weighted average distance (the dot size and circle size is indicative of the spread in reactor distance). The shaded region indicates the range of flux predictions corresponding to the 95% C.L. LMA region found in a global analysis of a solar neutrino data [Fogli *et al.*, 2002]. The thick curve corresponds to $\sin^2 2q_{12}$=0.83 and $\Delta m_{12}^2$ =2.5·10$^5$ eV$^2$. The dotted curve corresponds to $\sin^2 2q_{12}$=0.833 and $\Delta m_{12}^2$ =5.5·10$^5$ eV$^2$ [*Fogli et al.*, 2002] and is representative of recent best-fit LMA predictions while the dashed curve shows the case of small mixing angle (or no oscillation). Adapted from [*Eguchi et al.*, 2002; *Eguchi et al.*, 2003].

**Fig.10.** The ratio ($\Re = N_{\text{obs}}/N_{\text{expected}}$) of measured to expected flux in KamLAND-experiment at fixed angle of mixing ($\sin^2 2q_{12}$=0.83) but at the different mass squared differences. The insert: theoretical antineutrino spectrums in KamLAND experiment at at fixed angle of mixing ($\sin^2 2q_{12}$=0.83) and the different mass squared difference ($\Delta m_{12}^2 \approx (2 \div 4) \cdot 10^{-5}$ eV$^2$). Vertical line corresponds to the analysis threshold at 2.6 MeV. The green curve corresponding to theoretical antineutrino spectrum in KamLAND experiment at $\Delta m_{12}^2$ =2.5·10$^{-5}$ eV$^2$ is selected on two correlated signs (the spectrum shape and value of $\Re$=0.429) for the KamLAND experimental data description.

**Fig.11.** $\tilde{n}_e$ energy spectra in KamLAND. Main panel, experimental points (solid black dots with error bars) together the total expectation obtained in KamLAND experiment (dotted black line) [*Araki et al.,* 2005b] and presented paper (thick solid blue line). Also shown are expected neutrino spectrum (solid green line) from Japan's reactor, the expected neutrino spectrum from georeactor (red line), the expected signals from $^{238}U$ (dashed red line) and $^{232}Th$ (dashed green line) geoneutrinos, $^{13}C(\alpha,n)^{16}O$ reactions (dashed blue line) and accidentals (dashed black line). Inset, expected spectra obtained in KamLAND experiment (solid black line) [*Araki et al.,* 2005b] and presented paper (solid green line) extended to higher energy.

**Fig.12.** The theoretical (which takes into account the georeactor operation) reactor antineutrino spectrum calculated on the base of new data [*Araki et al.,* 2005] for all energy range of event detection. Designations are like in Fig.11.Vertical line corresponds to the analysis threshold at 2.6 MeV.

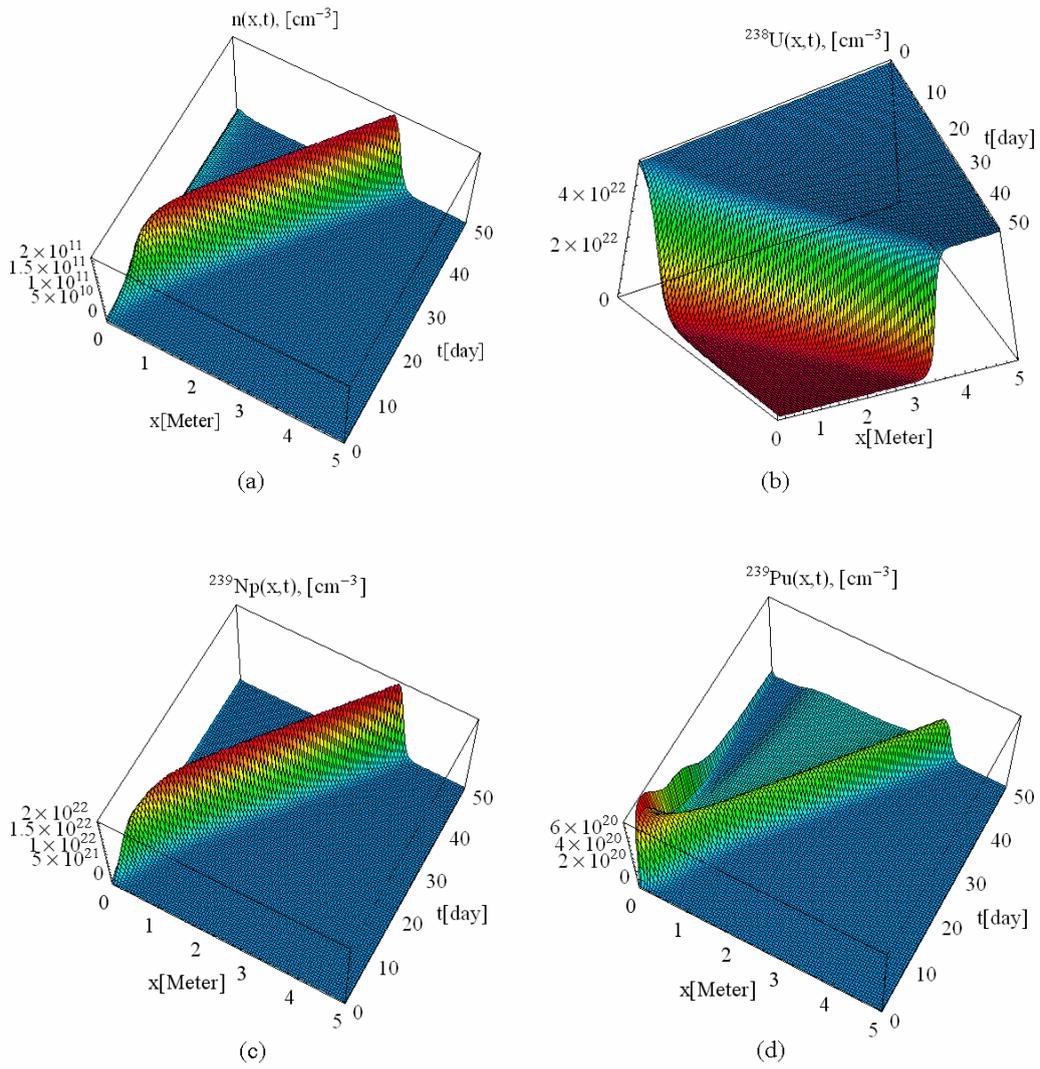

Fig.1.

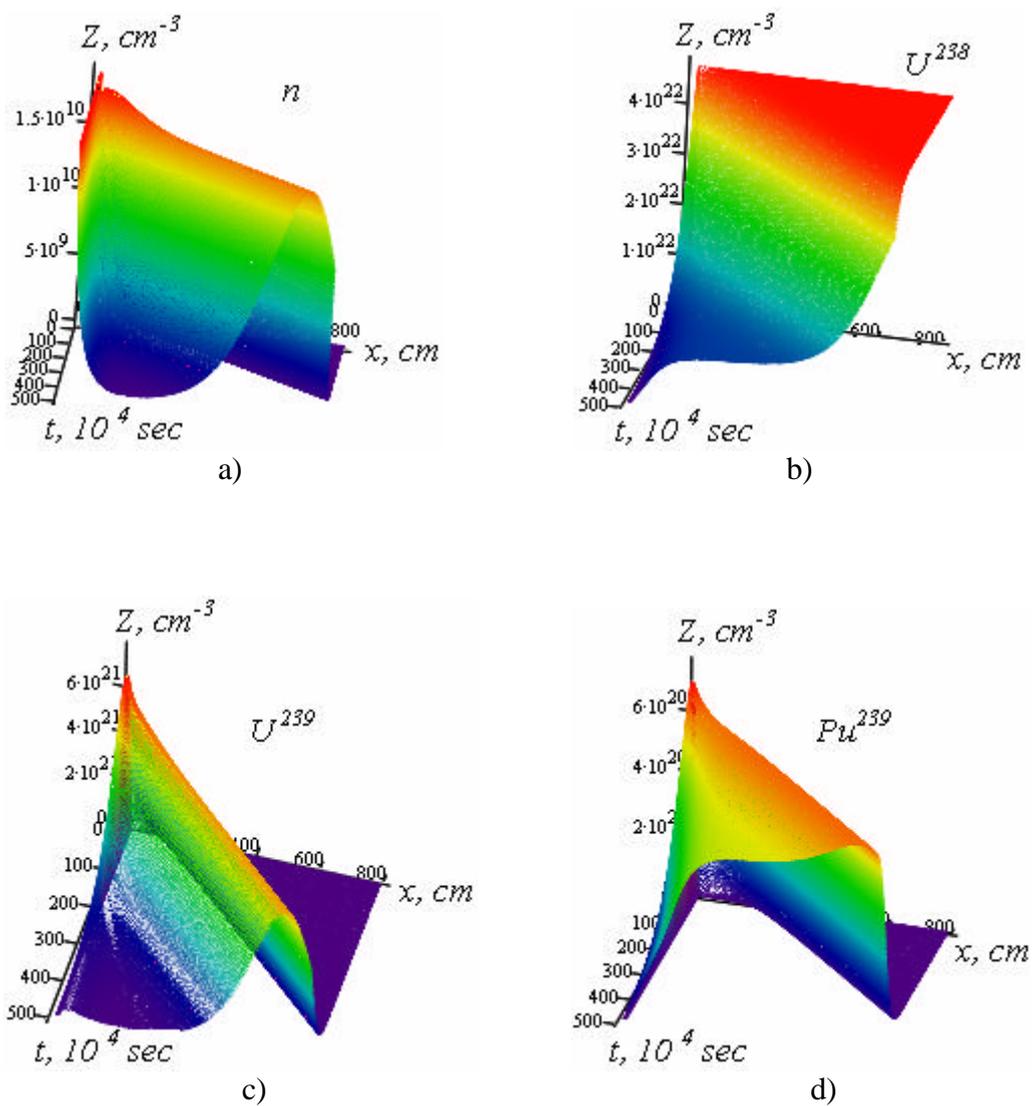

Fig. 2

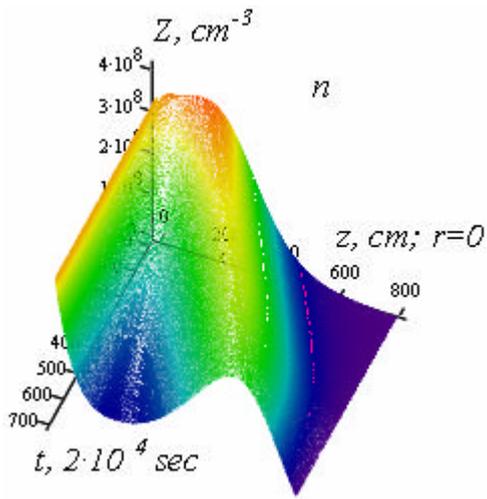

a)

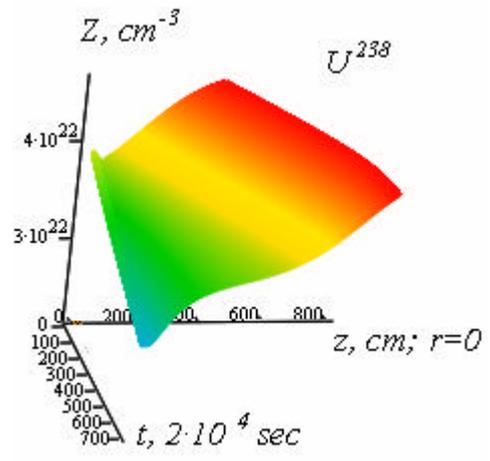

b)

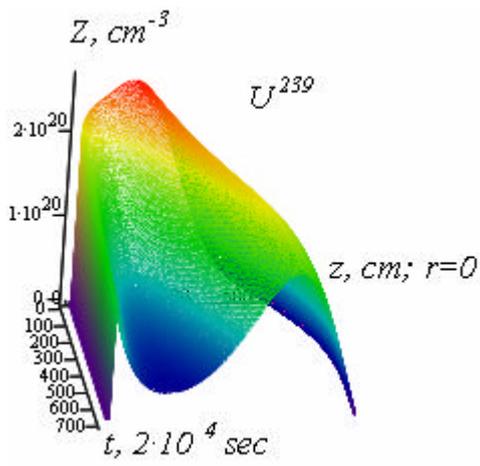

c)

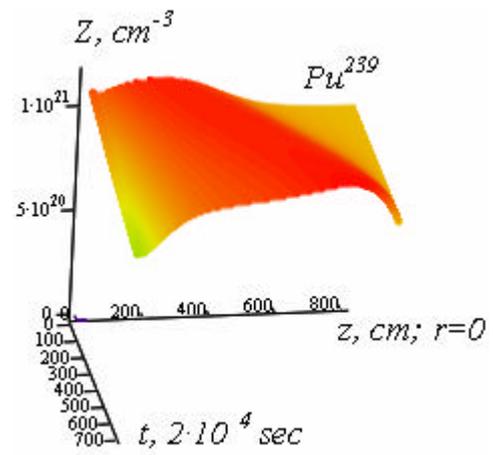

d)

Fig. 3

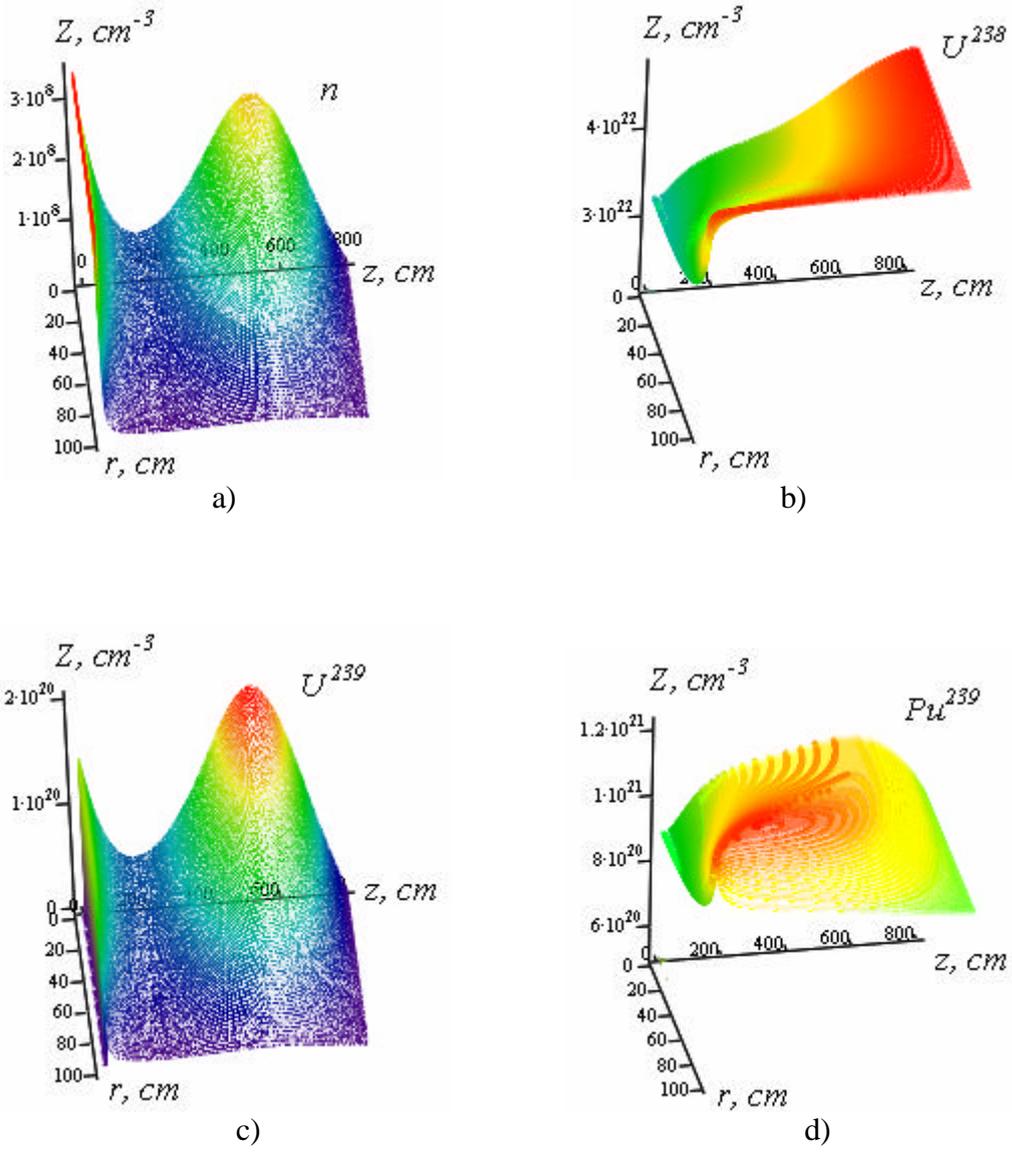

Fig .4

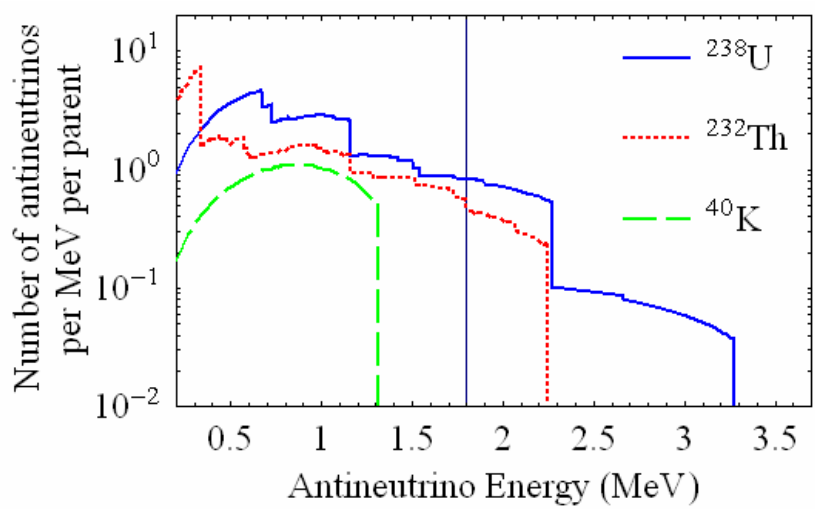

Fig. 5

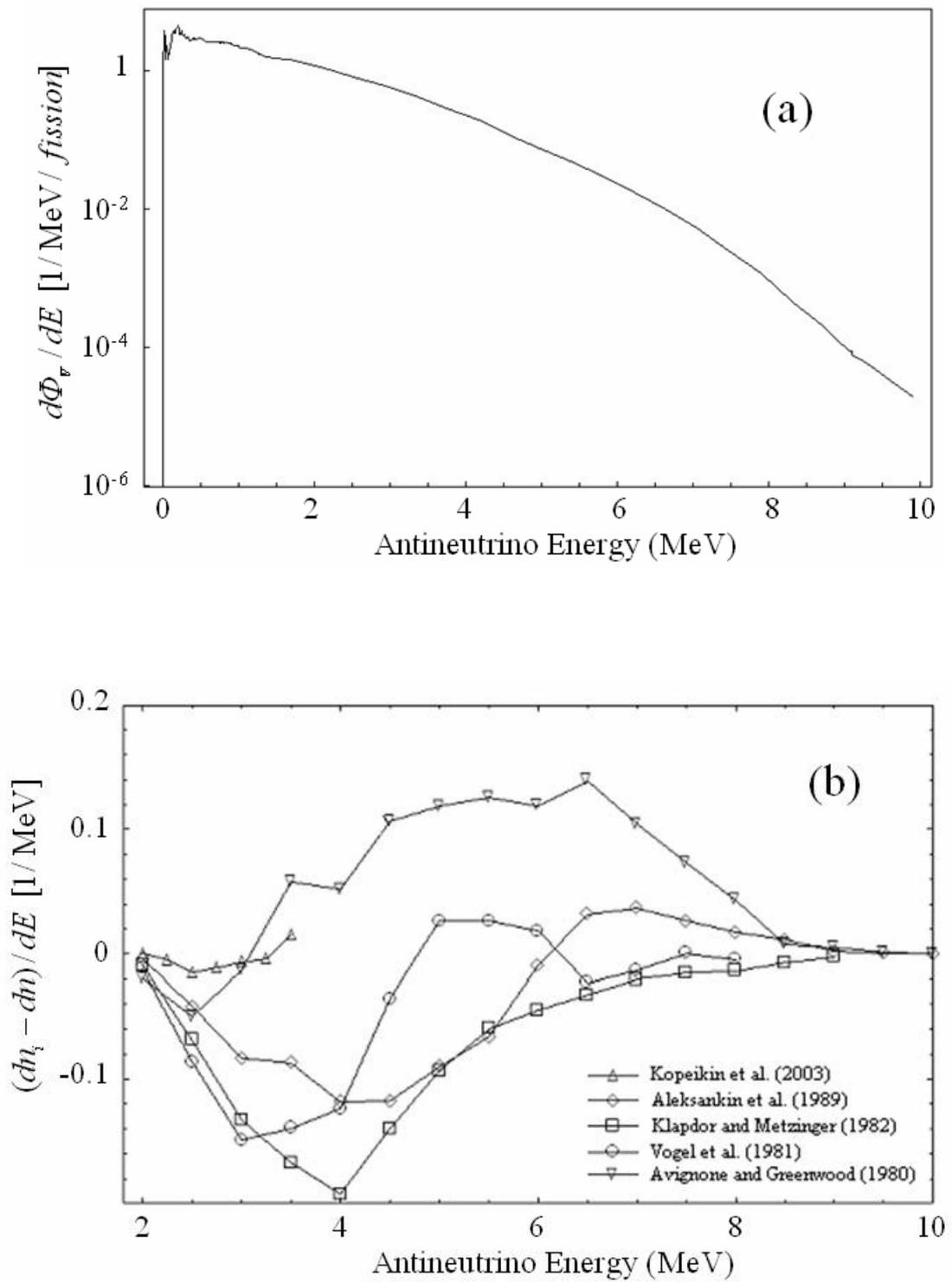

Fig. 6.

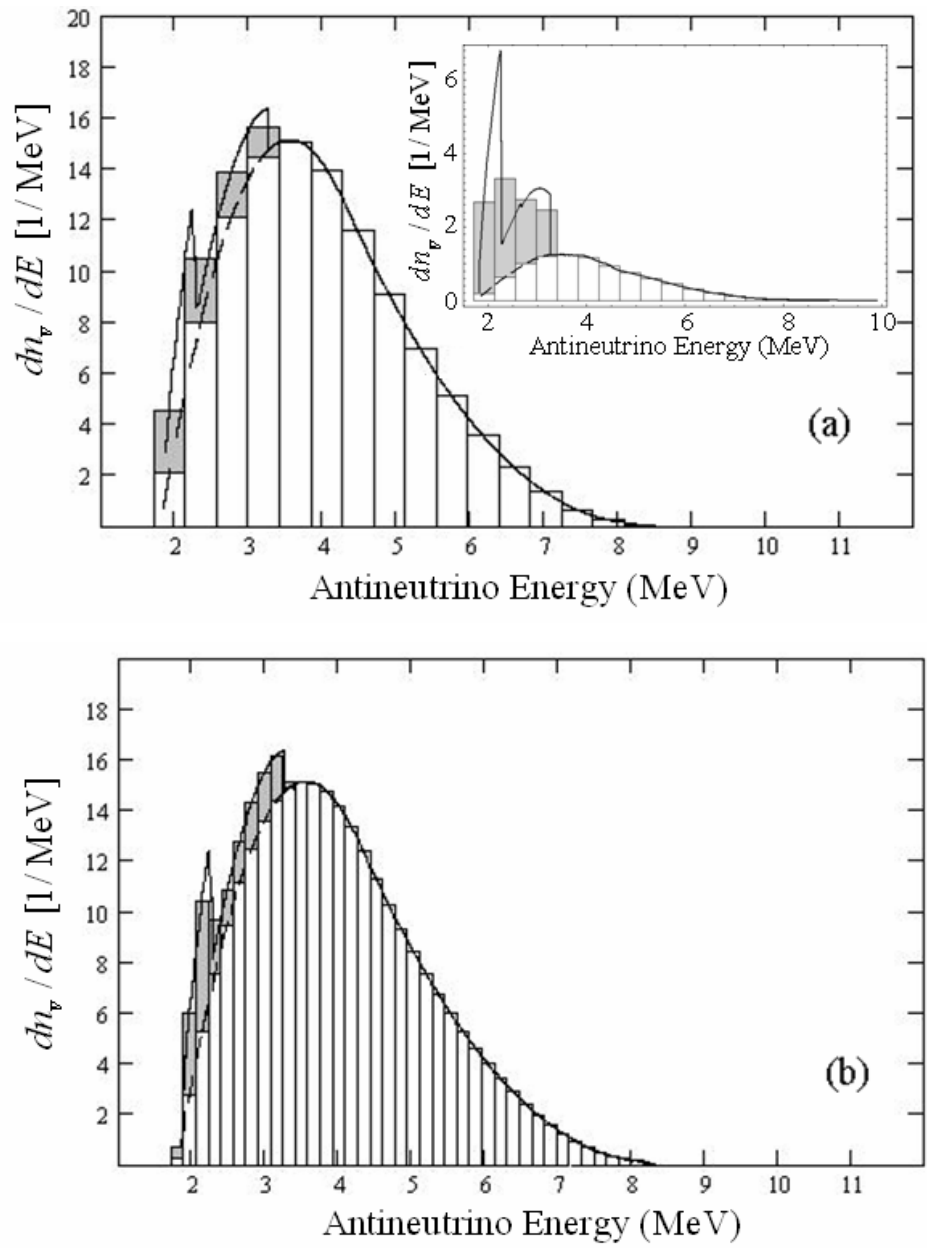

Fig. 7.

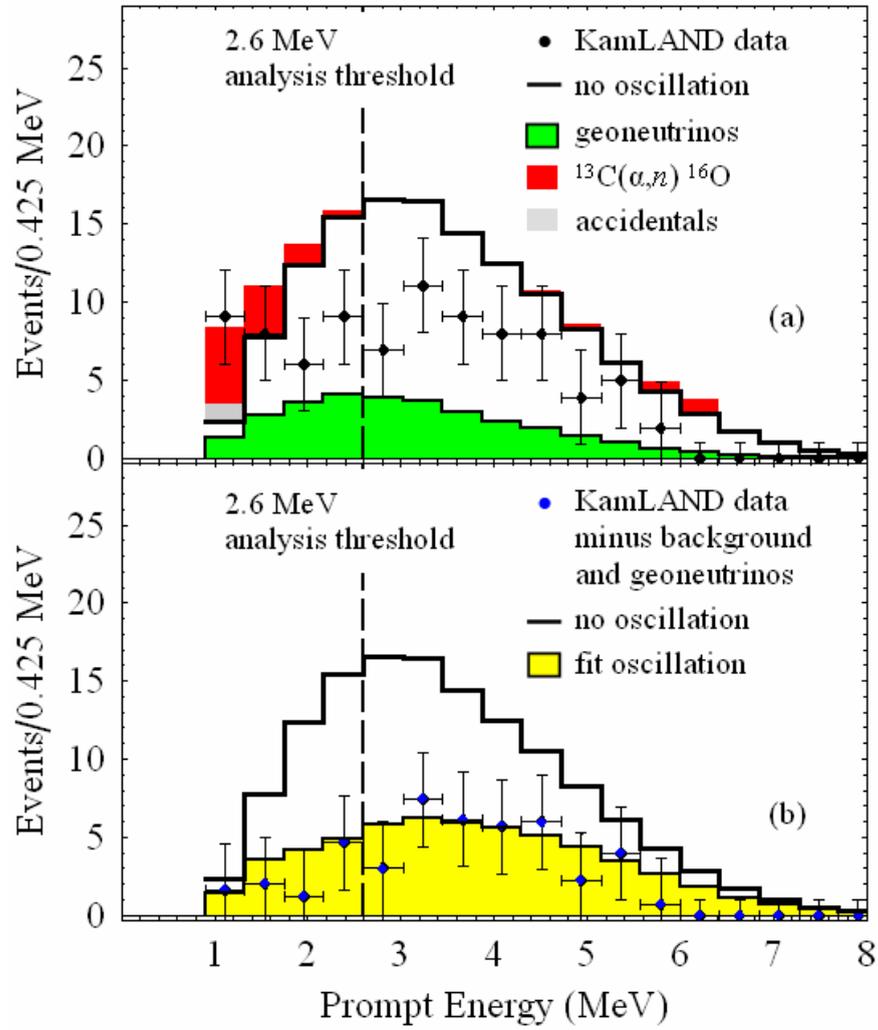

Fig. 8.

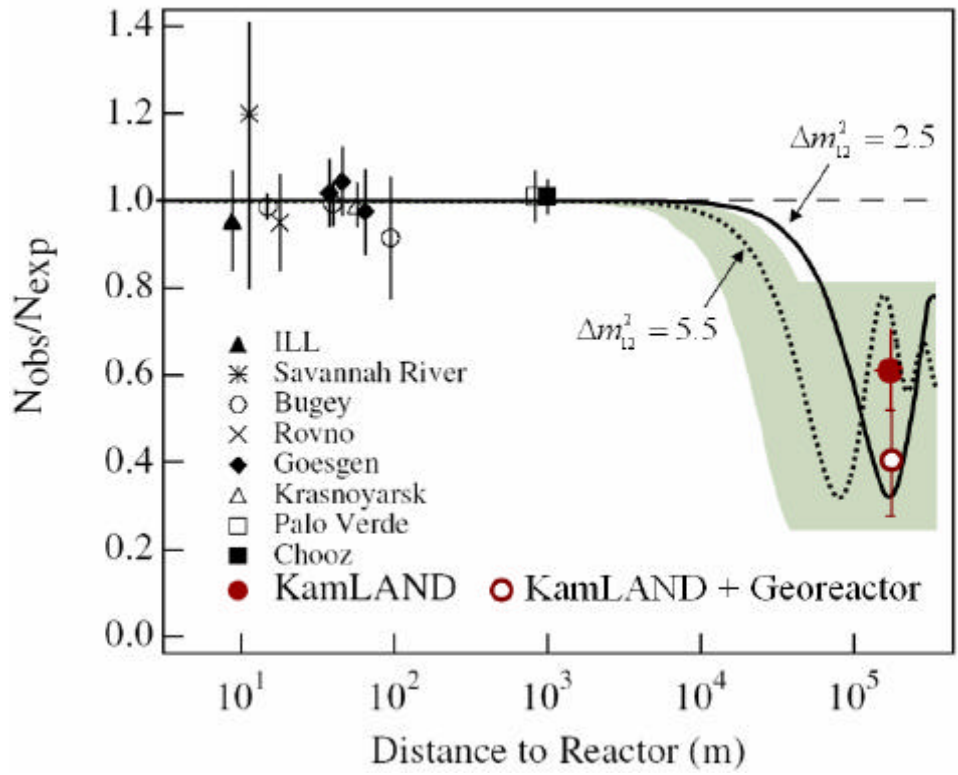

Fig. 9.

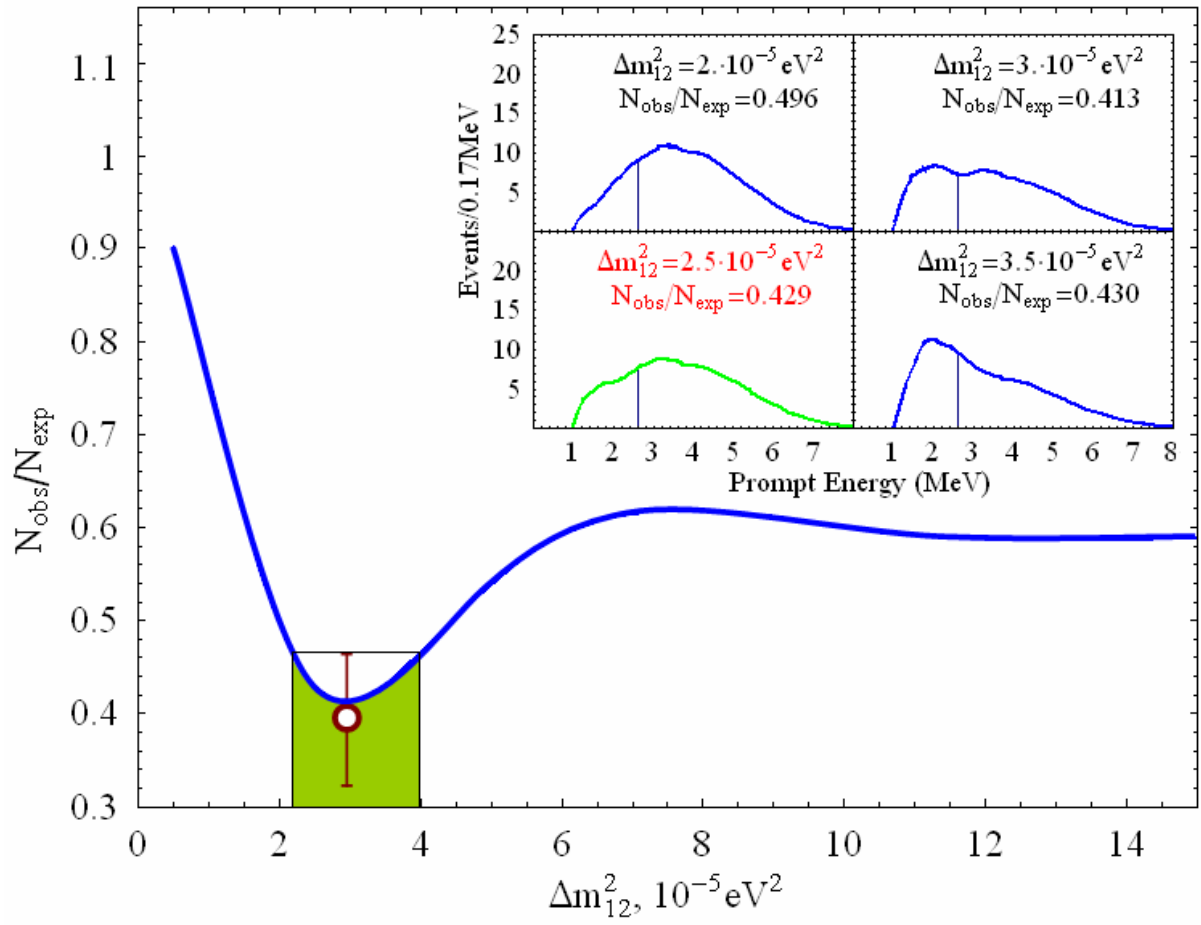

Fig. 10

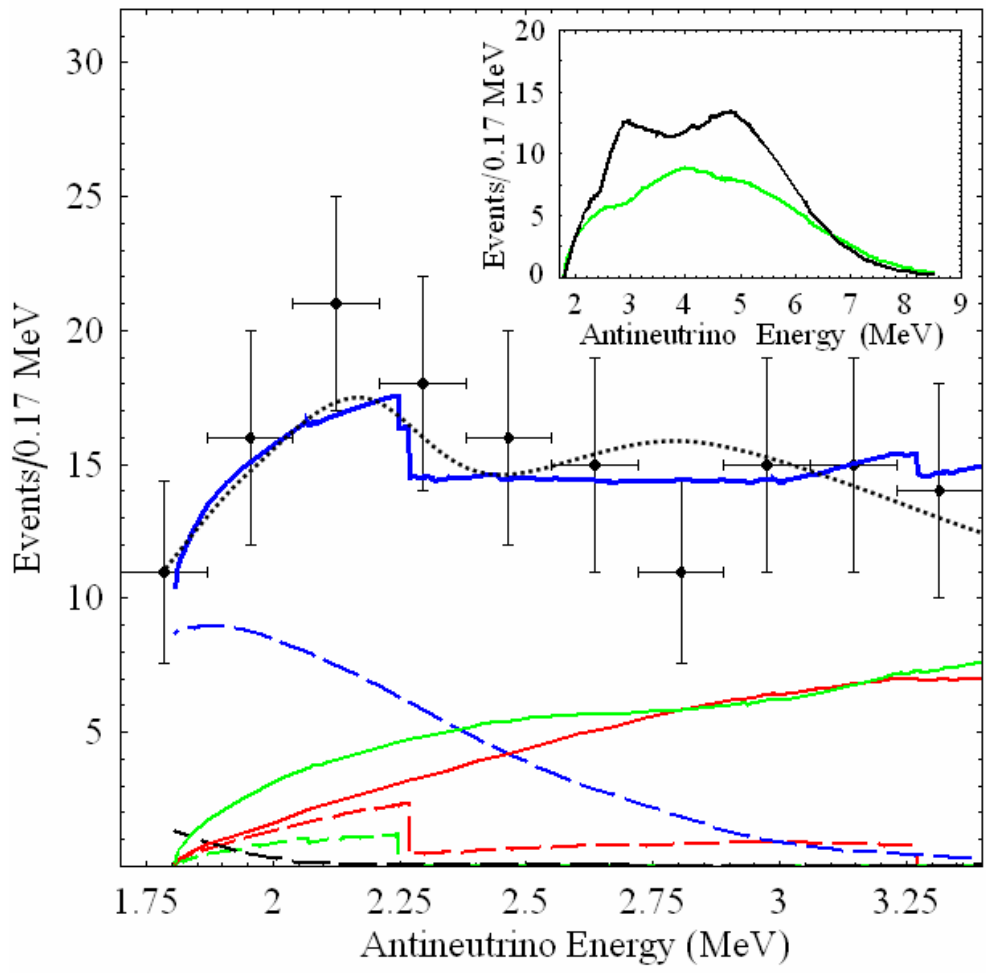

Fig.11.

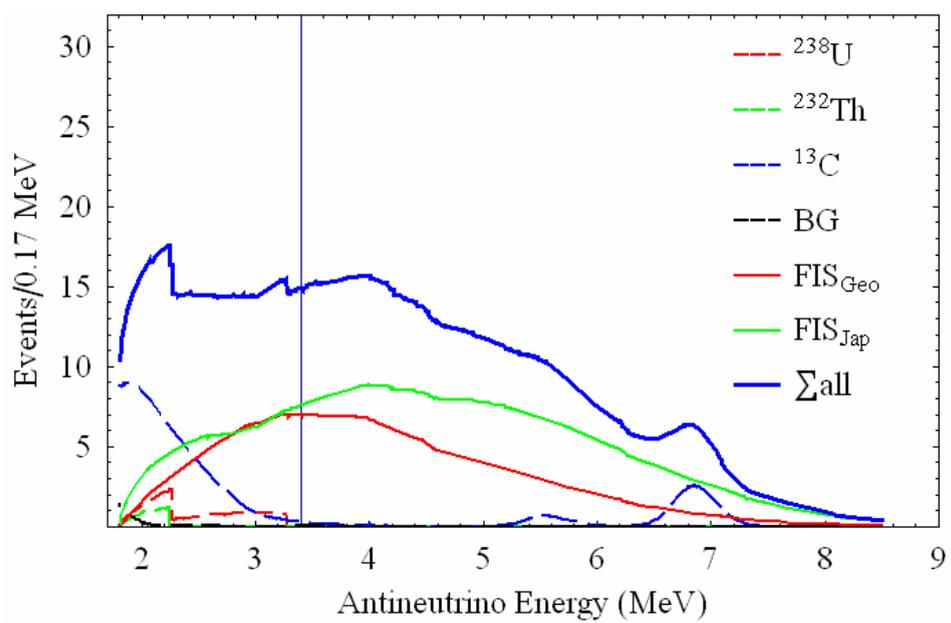

Fig. 12.